\documentclass[10pt,onecolumn,oneside,a4paper]{article}

\usepackage{amsmath,amssymb,amsfonts,amsthm}
\usepackage{authblk}
\usepackage{stackengine}
\usepackage{hyperref}
\usepackage[dvipsnames]{xcolor}
\allowdisplaybreaks

\theoremstyle{plain} \newtheorem{lemma}{\textbf{Lemma}}
\theoremstyle{plain} 
\theoremstyle{remark} \newtheorem{remark}{\textbf{Remark}}
\theoremstyle{plain} \newtheorem{theorem}{\textbf{Theorem}}
\theoremstyle{plain} \newtheorem{example}{\textbf{Example}}
\theoremstyle{plain} \newtheorem{corollary}{\textbf{Corollary}}
\theoremstyle{definition} 

\makeatletter
\newcommand{\pushright}[1]{\ifmeasuring@#1\else\omit\hfill$\displaystyle#1$\fi\ignorespaces}
\newcommand{\pushleft}[1]{\ifmeasuring@#1\else\omit$\displaystyle#1$\hfill\fi\ignorespaces}
\makeatother

\makeatletter
\let\@@pmod\pmod
\DeclareRobustCommand{\pmod}{\@ifstar\@pmods\@@pmod}
\def\@pmods#1{\mkern4mu({\operator@font mod}\mkern 6mu#1)}
\makeatother

\title{Feedback linearly extended discrete functions}
\author{Claude Gravel}
\affil{\stackunder{\small{EAGLYS Inc., Japan}}{\stackunder{\mbox{\small{\texttt{claudegravel1980@gmail.com}}}}{\mbox{\small{\texttt{c\_gravel@eaglys.co.jp}}}}}}
\author{Daniel Panario}
\affil{\stackunder{\small{School of Mathematics and Statistics}}{\stackunder{\small{Carleton University, Canada}}{\mbox{\small{\texttt{daniel@math.carleton.ca}}}}}}
\date{\today}

\begin{document}



\maketitle

\begin{abstract}
We study a new flexible method to extend linearly the graph of a non-linear, and usually not bijective, function so that the resulting extension is a bijection. Our motivation comes from cryptography. Examples from symmetric cryptography are given as how the extension was used implicitly in the construction of some well-known block ciphers. The method heavily relies on ideas brought from linear coding theory and secret sharing. We are interested in the behaviour of the composition of many extensions, and especially the space of parameters that defines a family of equations based on finite differences or linear forms. For any linear extension, we characterize entirely the space of parameters for which such equations are solvable in terms of the space of parameters that render those equations for the corresponding non-linear extended functions solvable. Conditions are derived to assess the solvability of those kind of equations in terms of the number of compositions or iterations. We prove a relation between the number of compositions and the dimensions of vector spaces that appear in our results. The proofs of those properties rely mostly on tools from linear algebra.


\textbf{AMS subject classifications:} 12E20 Finite fields; 15A03 Vector spaces, linear dependence, rank, lineability; 15B10 Orthogonal matrices; 39A06 Linear difference equations; 39A12 Discrete version of topics in analysis; 94A60 Cryptography; 94B05 Linear codes, general.

\end{abstract}


\section{Introduction}

For some integers $r,c>0$, let $\mathbf{I}_{r}$ and $\mathbf{0}_{r,c}$ denote the identity matrix of size $r\times r$ and the zero matrix of size $r\times c$, respectively, where the indices may be dropped whenever the sizes of the matrices are clear from the context.

Let $\mathbb{F}$ be a finite field. Let also $n>0$, $0<m<n$, be integers and let $f:\mathbb{F}^{m}\to \mathbb{F}^{n-m}$ be a given function. To define our mathematical object of interest, we need to consider two orthogonal subspaces of $\mathbb{F}^{n}$, each of dimension $m$ and $n-m$. That is we consider $\mathbb{F}^{n}=\mathbb{F}^{m}\oplus \mathbb{F}^{n-m}$. Let the matrices $\mathbf{A}$ and $\mathbf{B}$ be representations for bases of $\mathbb{F}^{m}$ and $\mathbb{F}^{n-m}$, respectively. The matrices $\mathbf{A}$ and $\mathbf{B}$ have size $m\times n$ and  $(n-m)\times n$, respectively. We can write $\mathbf{A}$ and $\mathbf{B}$ in standard form as $\mathbf{A}=[\mathbf{I}_{m}\mid \mathbf{C}]$ and $\mathbf{B}=[-\mathbf{C}^{{t}}\mid \mathbf{I}_{n-m}]$ for some matrix $\mathbf{C}$. We have by definition of orthogonality that $\mathbf{AB}^{{t}}=\mathbf{0}_{m,n-m}$. In addition, let $\mathbf{T}\in\mathrm{GL}(n,\mathbb{F})$, the linear group of dimension $n$ over $\mathbb{F}$. We shall be interested in the properties of the function $F$ such that
\begin{align}
&F:\mathbb{F}^{n}\to \mathbb{F}^{n}\nonumber\\
&x\mapsto \mathbf{T}(x+\mathbf{B}^{{t}}f(\mathbf{A}x)).\label{def_eq_rndfct}
\end{align}

We mention that a rich set of possible constructions for $\mathbf{A}$ and $\mathbf{B}$ arises from linear coding theory and error correction codes for which we refer to~\cite{Hill_1986}. For readers that are not acquainted with finite fields, we refer to~\cite{LidlNei1997}, and to~\cite{MullenPanario2013} for an exhaustive review of recent research in finite fields.

If we defined $F(x)=\mathbf{B}^{{t}}f(\mathbf{A}x)$ or $F(x)=\mathbf{TB}^{{t}}f(\mathbf{A}x)$, then $F$ would not be bijective, and so a justification is given for the feedback by $x$ in the defining Equation (\ref{def_eq_rndfct}) which is reminiscent to some non-linear feedback shift register as in \cite{Gol_2017_3rd}. The non-linear function $f$ can be chosen arbitrarily for the purpose of our work, but, from a practical point of view, $f$ is likely to be chosen uniformly and randomly from some family of functions. The function $F$ extends linearly the domain and image of $f$, and is even a bijection as we shall see soon. We study how some properties of $f$ are or are not transported into the linear extension $F$.

Throughout Section \ref{sect_results}, we shall point out to a non-exhaustive list of other research papers when necessary. Also we shall introduce concrete well-known examples when necessary. We recall a few facts or definitions, as in \cite{Godement} for instance, from linear algebra to end this section.

By convention, we assume that vectors are given in column format. Given a matrix $\mathbf{M}$ of size $r\times c$ and vector $x$, we can either multiply from the left or the right. If left multiplication is performed, then $x\in\mathbb{F}^{r}$, and we have $x^{t}\mathbf{M}$. If right multiplication is performed, then $x\in\mathbb{F}^{c}$, and we have $\mathbf{M}x$. For the left multiplication, we have a linear combination of the transposed rows of $\mathbf{M}$. For the right multiplication, we have a linear combination of the columns of $\mathbf{M}$. We have that
\begin{equation}
x^{t}\mathbf{M}=\sum_{i=1}^{r}{x_{i}(\mathrm{row}_{i}(\mathbf{M}))^{t}}\quad\text{and}\quad\mathbf{M}x=\sum_{i=1}^{c}{x_{i}\mathrm{col}_{i}(\mathbf{M})}.\label{fix_my_brain}
\end{equation}
A vector $z\in\mathrm{rowsp}\hspace{1pt}\mathbf{M}$ if and only if there exists $x\in\mathbb{F}^{r}$ such that $z^{t}=x^{t}\mathbf{M}$, or equivalently, $z=\mathbf{M}^{t}x$. A vector $z\in\mathrm{colsp}\hspace{1pt}\mathbf{M}$ if and only if there exists $x\in\mathbb{F}^{c}$ such that $z=\mathbf{M}x$, or equivalently, we might say that $z\in\mathrm{range}(\mathbf{M})$. Equations from (\ref{fix_my_brain}) are equivalent to assert that $\mathrm{colsp}\hspace{1pt}\mathbf{M}^{t}=\mathrm{rowsp}\hspace{1pt}\mathbf{M}$. The rank of a matrix is the number of linearly independent rows or columns, and sometimes the terms column rank or row rank are used in the literature. It holds that both column and row ranks are equal so that we can refer simply to the rank. Suppose the matrix $\mathbf{M}$ has rank $k\leq r$ and, without loss of generality, assume that $r\leq c$. We have that
\begin{displaymath}
\dim\mathrm{colsp}\hspace{1pt}\mathbf{M}=\dim\mathrm{rowsp}\hspace{1pt}\mathbf{M}=k,\hspace{2pt}\dim\ker\mathbf{M}=c-k,\hspace{2pt}\text{and}\hspace{2pt}\dim\ker\mathbf{M}^{t}=r-k.
\end{displaymath}
We have also that
\begin{displaymath}
\ker\mathbf{M}=(\mathrm{colsp}\hspace{1pt}\mathbf{M}^{t})^{\perp}=(\mathrm{rowsp}\hspace{1pt}\mathbf{M})^{\perp}\hspace{2pt}\text{and}\hspace{2pt}\ker\mathbf{M}^{t}=(\mathrm{colsp}\hspace{1pt}\mathbf{M})^{\perp}.
\end{displaymath}
We recall the Fredholm alternative theorem for the finite dimensional case which asserts that $\mathrm{colsp}\hspace{1pt}\mathbf{M}=(\ker\mathbf{M}^{t})^{\perp}$. Fredholm alternative theorem is equivalent to $\ker\mathbf{M}=(\mathrm{colsp}\hspace{1pt}\mathbf{M}^{t})^{\perp}$ applied on $\mathbf{M}^{t}$ in lieu of $\mathbf{M}$, and by using the fact the complement of the complement of a subspace is the subspace itself.

If $\mathbf{M}$ is a full-rank matrix of size $r\times c$, and without loss of generality $r\leq c$, then the canonical right projection of $\mathbf{M}$, denoted $\mathbf{R_{M}}$, is given by
\begin{equation}
\mathbf{R_{M}}=\left[
\begin{array}{c}
\mathbf{I}_{r} \\
\mathbf{0}_{c-r,r}
\end{array}\right].\label{def_rightinvstdform}
\end{equation}
If $\mathbf{M}$ is in standard form, that is $\mathbf{M}=[\mathbf{I}_{r}~\mathbf{M}']$ for some $r\times(c-r)$ matrix $\mathbf{M}'$, then
\begin{displaymath}
\mathbf{MR_{M}}=\mathbf{I}_{r}\quad\text{and}\quad\mathbf{R_{M}M}=\left(\begin{array}{c}\mathbf{M}\\  \mathbf{0}_{c-r,c}\end{array}\right).
\end{displaymath}

\section{Results}\label{sect_results}

We divide this section into three sub-sections for clarity. Section \ref{subsect_bij} contains a proof that the linear extension is a bijection even if the smaller extended function is not bijective. We find appropriate to include some well-known examples that are instances of our linear extension in Section \ref{subsect_bij}. In Section \ref{subsect_eqn_fd_lf}, we explore equations based on finite differences and linear forms, and we show how those equations can be solved using only the knowledge of the solutions for the smaller extended function. In Section \ref{subsect_comp}, we look upon generalizing Section \ref{subsect_eqn_fd_lf} to compositions of linear extensions.

\subsection{Bijective linear extension}\label{subsect_bij}

First, let us show the following lemma which uses the fact that $\mathbf{AB}^{{t}}=\mathbf{0}_{m,n-m}$.
\begin{lemma}
Let $\mathbb{F}$ be a finite field; the function $F$, as defined by Equation (\ref{def_eq_rndfct}), is bijective.
\end{lemma}
\begin{proof}
Because both the domain and the range of $F$ are equal and finite, we only need to show that $F$ is injective. Let $(x_1,y_1), (x_2,y_2)\in\mathbb{F}^{n}\times \mathbb{F}^{n}$ be such that $y_i=F(x_i)$ for $i=1,2$ and $y_2=y_1$. By using (\ref{def_eq_rndfct}), we have that
\begin{eqnarray*}
0=y_2-y_1
 &=& \mathbf{T}\Big(x_2+\mathbf{B}^{t}\big(f(\mathbf{A}x_{2})\big)\Big)
   - \mathbf{T}\Big(x_1+\mathbf{B}^{t}\big(f(\mathbf{A}x_{1})\big)\Big) \\
 &=& \mathbf{T}\Big(\big(x_2-x_1\big)+\mathbf{B}^{t}\big(f(\mathbf{A}x_{2})-f(\mathbf{A}x_1)\big)\Big),
\end{eqnarray*}
and therefore, since $\mathbf{T}$ is invertible,
$0=x_2-x_1+\mathbf{B}^{t}(f(\mathbf{A}x_{2})-f(\mathbf{A}x_1))$. Applying
$\mathbf{A}$ on the latter equation and using
that $\mathbf{AB}^{{t}}=\mathbf{0}_{m,n-m}$ entails $\mathbf{A}x_2=\mathbf{A}x_1$,
which implies that $f(\mathbf{A}x_2)=f(\mathbf{A}x_1)$. In turn, this leads to
$\mathbf{B}^{t}\big(f(\mathbf{A}x_2)-f(\mathbf{A}x_1)\big)=0$, and thus $x_1=x_2$.
\end{proof}

Let $x,y\in \mathbb{F}^{n}$ be such that $y=F(x)$. The inverse
$G$ of the permutation $F$ is given by
\begin{displaymath}
G(y)=\mathbf{T}^{-1}y-\mathbf{B}^{{t}}f(\mathbf{AT}^{-1}y).
\end{displaymath}
Indeed, using that $\mathbf{AB}^{{t}}=\mathbf{0}_{m,n-m}$, we have
\begin{align*}
GF(x)&=\mathbf{T}^{-1}F(x)-\mathbf{B}^{{t}}f(\mathbf{AT}^{-1}F(x))\\
&=\mathbf{T}^{-1}\big(\mathbf{T}(x+\mathbf{B}^{{t}}f(\mathbf{A}x))\big)-\mathbf{B}^{{t}}f(\mathbf{AT}^{-1}\mathbf{T}(x+\mathbf{B}^{{t}}f(\mathbf{A}x)))\\
&=x+\mathbf{B}^{{t}}f(\mathbf{A}x)-\mathbf{B}^{{t}}f(\mathbf{A}(x+\mathbf{B}^{{t}}f(\mathbf{A}x)))\\
&=x.
\end{align*}

We point out well-known examples from cryptography about how the linear extension from Equation (\ref{def_eq_rndfct}) has been used implicitly to construct block ciphers. Since we focus on the linear extension in this research, we do not explicitly recall the non-linear parts. Example \ref{ex_feistel} shows how our scheme encapsulates the original family of Feistel block ciphers. Example \ref{ex_idea_nxt} shows a more specific instance which is the FOX block cipher. As a companion of Examples \ref{exa_typeIfeistel} and \ref{exa_typeIIIfeistel}, we suggest the reader uses~\cite[Figure 1]{HoangR10} where the designs of several generalized Feistel networks are given.

\begin{example}[Feistel block cipher family]\label{ex_feistel}
Feistel block ciphers form a family of symmetric block ciphers, see \cite{HoangR10}. We show in this example that they can be seen as a subfamily of our scheme. The non-linear part is taken from a certain family of functions. The working field is $\mathbb{F}_{2}$ that we denote by $\mathbb{F}$ for this example. Each non-linear function is indexed by a key denoted by $k$ which is then used to index the linear extension. With the values $m=1$ and $n=2m$, the non-linear $f_k$ can be specified to design particular instances of a Feistel network such as DES. An input $x$ is a two-block column vector given by
\begin{displaymath}
x=\left(\begin{array}{c}
x_0\\
x_1\end{array}\right)
\end{displaymath}
with $x_0,x_1\in \mathbb{F}^{d}$ for some $d>1$, and similarly for an output vector. The non-linear function $f_k$ is defined over $\mathbb{F}^{d}$. If $\mathbf{I}=\mathbf{I}_{d}$ and $\mathbf{0}=\mathbf{0}_{d,d}$, then the linear transformations $\mathbf{A}$, $\mathbf{B}$, and $\mathbf{T}$ are respectively given by
\begin{displaymath}
\mathbf{A}=\left(\begin{array}{cc}
\mathbf{I} & \mathbf{0}
\end{array}\right),\quad \mathbf{B}=\left(\begin{array}{cc}
\mathbf{0} & \mathbf{I}
\end{array}\right),\quad\text{and}\quad \mathbf{T}=\left(\begin{array}{cc}
\mathbf{0} & \mathbf{I}\\
\mathbf{I} & \mathbf{0}
\end{array}\right).
\end{displaymath}
Finally we have that
\begin{align*}
F_{k}(x)&=F_{k}\left(\begin{array}{c}
x_0\\
x_1\end{array}\right)=\mathbf{T}\big(x+\mathbf{B}^{t}f_{k}(\mathbf{A}x)\big)\\
&=\left(\begin{array}{cc}
\mathbf{0} & \mathbf{I}\\
\mathbf{I} & \mathbf{0}
\end{array}\right)\Bigg(\left(\begin{array}{c}
x_0\\
x_1\end{array}\right)+\left(\begin{array}{c}
\mathbf{0}\\
\mathbf{I}\end{array}\right)f_{k}\bigg(\left(\begin{array}{cc}
\mathbf{I} & \mathbf{0}
\end{array}\right)\left(\begin{array}{c}
x_0\\
x_1\end{array}\right)\bigg)\Bigg)\\
&=\left(\begin{array}{c}
x_1+f_{k}(x_0)\\
x_0\end{array}\right).
\end{align*}

\hfill$\square$
\end{example}

We observe the matrix $\mathbf{T}$ from Example \ref{ex_feistel} is a permutation. Another example is IDEA NXT that uses the Lai-Massey scheme \cite{Lai:1991:PNB:112331.112375} as a building primitive.
\begin{example}[IDEA NXT\---FOX]\label{ex_idea_nxt}
The working field is $\mathbb{F}_{2}$ as in the previous example. Here $d=16$ or $d=32$, $m=2$, and $n=2m$. An input $x$ is written column-wise as four consecutive blocks $L_0$, $L_1$, $R_0$, $R_1\in\mathbb{F}^{d}=\mathbb{F}^{d\times 1}$, that is,
\begin{align*}
x&=\left(\begin{array}{c}L_0\\L_1\\R_0\\R_1\end{array}\right).
\end{align*}
We denote $\mathbf{I}_{d}$ and $\mathbf{0}_{d,d}$ by $\mathbf{I}$ and $\mathbf{0}$, respectively. The matrices $\mathbf{A}$, $\mathbf{B}$, and $\mathbf{T}$ are given by
\begin{displaymath}
\mathbf{A}=\left(
\begin{array}{cccc}
\mathbf{I} & \mathbf{0} & \mathbf{I} & \mathbf{0}\\
\mathbf{0} & \mathbf{I} & \mathbf{0} & \mathbf{I}
\end{array} \right),\quad
\mathbf{B}=\mathbf{A},\quad
\mathbf{T}=\left(
\begin{array}{cccc}
\mathbf{0} & \mathbf{I} & \mathbf{0} & \mathbf{0}\\
\mathbf{I} & \mathbf{I} & \mathbf{0} & \mathbf{0}\\
\mathbf{0} & \mathbf{0} & \mathbf{I} & \mathbf{0}\\
\mathbf{0} & \mathbf{0} & \mathbf{0} & \mathbf{I}
\end{array}
\right).
\end{displaymath}

The non-linear keyed function $f_{k}:\mathbb{F}^{d}\times\mathbb{F}^{d}\mapsto \mathbb{F}^{d}\times\mathbb{F}^{d}$ is as given in \cite{FOX} with $k\in \mathbb{F}^{d}\times\mathbb{F}^{d}\times\mathbb{F}^{d}\times\mathbb{F}^{d}$. Let $(z_0,z_1)^{{t}}=f_{k}(y_0,y_1)^{{t}}$  for $z_0$, $z_1$, $y_0$, and $y_1\in \mathbb{F}^{d}$. Given an input $x=(L_0,L_1,R_0,R_1)^{{t}}$ and a round key $k\in\mathbb{F}^{d}\times\mathbb{F}^{d}\times\mathbb{F}^{d}\times\mathbb{F}^{d}$, one round of FOX is given by
\begin{align*}
F_{k}(x)&=\mathbf{T}\big(x+\mathbf{B}^{{t}}f_{k}(\mathbf{A}x)\big)=\mathbf{T}\left(x+\mathbf{B}^{{t}}f_{k}\left(\begin{array}{c}L_0+R_0\\L_1+R_1\end{array}\right)\right)\\
&=\mathbf{T}\left(\left(\begin{array}{c}L_0\\L_1\\R_0\\R_1\end{array}\right)+\left(\begin{array}{cc}\mathbf{I} & \mathbf{0}\\ \mathbf{0} & \mathbf{I}\\ \mathbf{I} & \mathbf{0} \\ \mathbf{0} & \mathbf{I}\end{array}\right)\left(\begin{array}{c}z_0\\z_1\end{array}\right)\right)=\left(\begin{array}{c}z_1+L_1\\z_0+z_1+L_0+L_1\\z_0+R_0\\z_1+R_1\end{array}\right).
\end{align*}
\hfill$\square$
\end{example}

\begin{example}[Type-$1$ generalized Feistel network]\label{exa_typeIfeistel}
In this example, the working field is $\mathbb{F}_{2^d}$ for some $d>1$ that we denote by $\mathbb{F}$. We have $m=1$, $n=4m$, $\mathbf{I}\in\mathbb{F}_{2}^{d\times d}$, $\mathbf{0}\in\mathbb{F}_{2}^{d\times d}$. An input $x$ is written column-wise as four consecutive blocks $x_i\in\mathbb{F}$ for $1\leq i\leq 4$. The non-linear part $f$ is defined over $\mathbb{F}$. (We note that in \cite{HoangR10}, capital $F$ denotes
a smaller non-linear function which is denoted $f$ by us. Also, blocks are denoted by $B_i$ and their length by $n$ in \cite{HoangR10}. The length of a block here is $d$.) The template matrices $\mathbf{A}\in\mathbb{F}^{1\times 4}$, $\mathbf{B}\in\mathbb{F}^{1\times 4}$ and $\mathbf{T}\in\mathbb{F}^{4\times 4}$ are given by
\begin{displaymath}
\mathbf{A}=\left(\begin{array}{cccc}
\mathbf{I} & \mathbf{0} & \mathbf{0} & \mathbf{0}
\end{array}\right),\quad
\mathbf{B}=\left(\begin{array}{cccc}
\mathbf{0} & \mathbf{I} & \mathbf{0} & \mathbf{0}
\end{array}\right),\quad
\mathbf{T}=\left(\begin{array}{cccc}
\mathbf{0} & \mathbf{I} & \mathbf{0} & \mathbf{0}\\
\mathbf{0} & \mathbf{0} & \mathbf{I} & \mathbf{0}\\
\mathbf{0} & \mathbf{0} & \mathbf{0} & \mathbf{I}\\
\mathbf{I} & \mathbf{0} & \mathbf{0} & \mathbf{0}
\end{array}\right).
\end{displaymath}
\hfill$\square$
\end{example}

From \cite{HoangR10}, alternating Feistel and unbalanced Feistel can be
also embedded in our scheme with minor changes. For alternating Feistel,
we use two instances of $\mathbf{A}$, $\mathbf{B}$, $\mathbf{T}$ and $f$. Our scheme allows $f$ to
be non-invertible and hence include unbalanced Feistel.

It is interesting that Type-$2$ and Type-$3$ generalized Feistel networks
\cite{HoangR10} do not fit in our model as presented so far. However, we can
expand our scheme so to include such networks. For the sake of conciseness, we show
this only for Type-$3$ generalized Feistel networks; Type-$2$ can be
easily derived by simplifying the Type-$3$ model.

\begin{example}[Type-$3$ generalized Feistel network]\label{exa_typeIIIfeistel}
The working field is as in the previous example. We have $m=1$, $n=4m$, $\mathbf{I}\in\mathbb{F}_{2}^{d\times d}$, $\mathbf{0}\in\mathbb{F}_{2}^{d\times d}$. An input $x$ is written column-wise as four consecutive blocks $x_i\in\mathbb{F}$ for $1\leq i\leq 4$. There are three non-linear functions $f_{j}:\mathbb{F}\to \mathbb{F}$ for $j=1,2,3$. The template matrices $\mathbf{A}_j\in\mathbb{F}^{1\times 4}$ and $\mathbf{B}_j\in\mathbb{F}^{1\times 4}$ are given by
\begin{align*}
\mathbf{A}_{1}=\left(\begin{array}{cccccccccccc}
\mathbf{I} & \mathbf{0} & \mathbf{0} & \mathbf{0}
\end{array}\right),\quad &
\mathbf{B}_{1}=\left(\begin{array}{cccccccccccc}
\mathbf{0} & \mathbf{I} & \mathbf{0} & \mathbf{0}
\end{array}\right),\\
\mathbf{A}_{2}=\left(\begin{array}{cccccccccccc}
\mathbf{0} & \mathbf{I} & \mathbf{0} & \mathbf{0}
\end{array}\right),\quad &
\mathbf{B}_{2}=\left(\begin{array}{cccccccccccc}
\mathbf{0} & \mathbf{0} & \mathbf{I} & \mathbf{0}
\end{array}\right),\\
\mathbf{A}_{3}=\left(\begin{array}{cccccccccccc}
\mathbf{0} & \mathbf{0} & \mathbf{I} & \mathbf{0}
\end{array}\right),\quad &
\mathbf{B}_{3}=\left(\begin{array}{cccccccccccc}
\mathbf{0} & \mathbf{0} & \mathbf{0} & \mathbf{I}
\end{array}\right).
\end{align*}
The matrix $\mathbf{T}$ is as in Example \ref{exa_typeIfeistel}. Finally, we have
\begin{align}
F_{k}\left(\begin{array}{c}x_1\\ x_2 \\ x_3\\ x_4\end{array}\right)
&=\mathbf{T}\left(\left(\begin{array}{c}x_1\\ x_2 \\ x_3\\ x_4\end{array}\right)
+\sum_{j=1}^{3}{\mathbf{B}^{\textrm{t}}_{j}f_{j}\left(\mathbf{A}_{j}
\left(\begin{array}{c}x_1\\ x_2 \\ x_3\\ x_4\end{array}\right)\right)}\right).
\label{typethree}
\end{align}
\hfill$\square$
\end{example}

We hope that the examples above are sufficient to serve as a justification why it matters to study the properties of linear extensions of the kind given by Equation (\ref{def_eq_rndfct}). An interesting way to construct $\mathbf{A}$, $\mathbf{B}$ is obviously by using the theory of linear codes. For that, we point out that self-dual linear codes such as the maximum distance separable codes, hereafter abbreviated by MDS, seem to result in extensions with interesting algebraic, combinatorial and statistical properties that we are currently investigating such as the cycle structure as in \cite{GraPanTho_UniPer_2019} and pattern distribution. Among the most important linear MDS codes are the Reed-Solomon and BCH codes that were invented independently by \cite{Bose_1960} and \cite{Hoc_1959}. We recall that Reed-Solomon codes are mathematically equivalent to Shamir’s secret sharing \cite{Sha_1979} as explained in \cite{McESar_1981}.

\subsection{Equations with finite differences and linear forms}\label{subsect_eqn_fd_lf}

We are interested to solve or search for solutions to equations involving finite differences or linear forms. More precisely, we want to solve for an equation of the form that is given either by $F(x+\alpha)-F(x)=\beta$ or by $\alpha\cdot x-\beta\cdot F(x)=0$ for some $\alpha,\beta\in \mathbb{F}^{n}$. The parameters defining the former equations are $\alpha$ and $\beta$, and the symbol $\cdot$ stands for the standard inner product. We recall that given $u,v\in\mathbb{F}^{n}$, then $u^{t}\in\mathbb{F}^{1\times n}$ is the transposition of $u$, and, by definition of the standard inner product, we have $u\cdot v=u^{t}v\in\mathbb{F}$. The last two equations are of particular importance in cryptography as it can be noticed from the following non-exhaustive list of documents: \cite{Biham1991}, \cite{Carlet1}, \cite{Carlet2}, \cite{Heys2002}, \cite{Matsui1994}, \cite{Sigenthaler1984}, or \cite{XiaoM88}.

We mention that dimensions of the fundamental subspaces associated to the matrices $\mathbf{A}$ and $\mathbf{B}$ are connected together. If the dimensions are taken over the finite field $\mathbb{F}$, then we have
\begin{equation}
m=\mathrm{dim}\big(\mathrm{rowsp}\hspace{1pt}\mathbf{A}\big)=\mathrm{dim}\big(\ker\mathbf{B}\big),~{}n-m=\mathrm{dim}\big(\ker\mathbf{A}\big)=\mathrm{dim}\big(\mathrm{rowsp}\hspace{1pt}\mathbf{B}\big).\label{big_smelly_gases}
\end{equation}
Because $\mathbf{AB}^{t}=\mathbf{0}$, then (\ref{big_smelly_gases}) just above can be restated as $\mathrm{rowsp}\hspace{1pt}\mathbf{A}=\ker\mathbf{B}$ and $\ker\mathbf{A}=\mathrm{rowsp}\hspace{1pt}\mathbf{B}$.

In our case, we are interested first to fix the matrices $\mathbf{A}$, $\mathbf{B}$, and $\mathbf{T}$, second to select a set of functions $\{f_{i}:\mathbb{F}^{m}\to\mathbb{F}^{n-m}\colon 1\leq i\leq \ell\}$, and third to solve for
\begin{align}
F_{\ell}\circ F_{\ell-1}\circ \cdots \circ F_{1}(x+\alpha)-F_{\ell}\circ F_{\ell-1}\circ \cdots \circ F_{1}(x)&=\beta\quad\text{or}\nonumber\\
\beta\cdot F_{\ell}\circ F_{\ell-1}\circ \cdots \circ F_{1}(x)&=\alpha\cdot x,\label{multiple_iter_def_eq}
\end{align}
where $F_{i}(x)=\mathbf{T}(x+\mathbf{B}^{{t}}f_{i}(\mathbf{A}x))$ for $1\leq i\leq \ell$. In practice, it might be very difficult to solve the preceding equations even for moderate values of $\ell$. Any conceivable ways to learn any information, probabilistically or deterministically, about the solution sets for a given pair $(\alpha,\beta)$ matter. \emph{And} what are those pairs $(\alpha,\beta)$ that render equations from (\ref{multiple_iter_def_eq}) feasible in $x\in\mathbb{F}^{n}$? When $\ell=1$ in (\ref{multiple_iter_def_eq}), we can characterize the solutions of $F(x+\alpha)-F(x)=\beta$ or $\alpha\cdot x-\beta\cdot F(x)=0$ solely in terms of the solutions of $f(u+a)-f(u)=b$ or $a\cdot u - b\cdot f(u)=0$, respectively, for some $u\in\mathbb{F}^{m}$, $a\in\mathbb{F}^{m}$ and $b\in\mathbb{F}^{n-m}$ to be specified hereafter in Lemmas \ref{lemma_fin_diff} and \ref{lemma_lin_approx}.

\begin{lemma}\label{lemma_fin_diff}
Given $\alpha,\beta\in\mathbb{F}^{n}$ such that $T^{-1}\beta - \alpha \in \mathrm{rowsp}\hspace{1pt}{B}$, the solution space for $F(x+\alpha)-F(x)=\beta$ is given by
\begin{displaymath}
\big\{x\in\mathbb{F}^{n}\colon\hspace{2pt}u=\mathbf{A}x,\hspace{2pt}a=\mathbf{A\alpha},\hspace{2pt}b=\mathbf{R_{B}^{{\text{t}}}}(\mathbf{T}^{-1}\beta-\alpha),\hspace{2pt}f(u+a)-f(u)=b\hspace{2pt}\big\},
\end{displaymath}
where $\mathbf{R_{B}}$ is defined in (\ref{def_rightinvstdform}).
\end{lemma}
\begin{proof}
Suppose $F(x+\alpha)-F(x)=\beta$ for given $\alpha,\beta$; then
\begin{align*}
\beta&=F(x+\alpha)-F(x)\\
&=\mathbf{T}\big(x+\alpha+\mathbf{B}^{{t}}f(\mathbf{A}(x+\alpha))\big)-\mathbf{T}\big(x+\mathbf{B}^{{t}}f(\mathbf{A}x)\big)\\
&=\mathbf{T}\alpha+\mathbf{TB}^{{t}}f(\mathbf{A}(x+\alpha))-\mathbf{TB}^{{t}}f(\mathbf{A}x)\\
&=\mathbf{T}\alpha+\mathbf{TB}^{{t}}\big(f(\mathbf{A}x+\mathbf{A}\alpha)-f(\mathbf{A}x)\big),
\end{align*}
which is equivalent to
\begin{equation}
\mathbf{B}^{{t}}\big(f(\mathbf{A}x+\mathbf{A}\alpha)-f(\mathbf{A}x)\big)=\mathbf{T}^{-1}\beta-\alpha.\label{eq_fin_dif_zoo}\\
\end{equation}
By using the canonical right projection of $\mathbf{B}$, then the previous equation is equivalent to
\begin{displaymath}
\mathbf{R_{B}^{\text{t}}\mathbf{B}^{{t}}}\big(f(\mathbf{A}x+\mathbf{A}\alpha)- f(\mathbf{A}x)\big)=\mathbf{R_{B}^{\text{t}}}\big(\mathbf{T}^{-1}\beta-\alpha\big)=f(\mathbf{A}x+\mathbf{A}\alpha)- f(\mathbf{A}x).
\end{displaymath}
The proof is complete by letting $u=\mathbf{A}x$, $a=\mathbf{A}\alpha$, and $b=\mathbf{R_{B}^{\text{t}}}\big(\mathbf{T}^{-1}\beta-\alpha\big)$.
\end{proof}

Lemma \ref{lemma_fin_diff} stipulates that the set of parameters (and hence the solution space) for the linear extension reduces linearly to the smaller set of parameters that define the equations for the extended function solely. In Lemma \ref{lemma_lin_approx} that follows immediately, we show that this is also the case for an equation involving linear forms.

\begin{lemma}\label{lemma_lin_approx}
Given $\alpha,\beta\in\mathbb{F}^{n}$ such that $\alpha^t - \beta^t T \in \mathrm{rowsp}\hspace{1pt}{A}$, the solution space for $\alpha\cdot x -\beta\cdot F(x)=0$ is given by
\begin{displaymath}
\big\{x\in\mathbb{F}^{n}\colon u=\mathbf{A}x,\hspace{2pt} a=\mathbf{R_{A}^{\text{t}}}\big(\alpha-\mathbf{T}^{{t}}\beta\big),\hspace{2pt}b=\beta^{t}\mathbf{T}\mathbf{B}^{{t}},\hspace{2pt}a\cdot u-b\cdot f(u)=0\big\},
\end{displaymath}
where $\mathbf{R_{A}}$ is defined in (\ref{def_rightinvstdform}).
\end{lemma}

\begin{proof}
Suppose $\alpha\cdot x -\beta\cdot F(x)=0$ for given $\alpha,\beta$; then
\begin{align}
0&=\alpha\cdot x-\beta\cdot F(x)\nonumber\\
&=\alpha\cdot x-\beta\cdot\big(\mathbf{T}(x+\mathbf{B}^{{t}}f(\mathbf{A}x))\big)\nonumber\\
&=\alpha^{t}x-\beta^{{t}}\big(\mathbf{T}(x+\mathbf{B}^{{t}}f(\mathbf{A}x))\big)\nonumber\\
&=\big(\alpha^{{t}}-\beta^{{t}}\mathbf{T}\big)x-\beta^{{t}}\mathbf{TB}^{{t}}f(\mathbf{A}x).\label{cocoboule}
\end{align}
By definition, $\alpha^{t}-\beta^{t}\mathbf{T}\in\mathrm{rowsp}\hspace{1pt}\mathbf{A}$ if and only if there is non-zero $a\in\mathbb{F}^{m}$ such that $a^{t}\mathbf{A}=\alpha^{t}-\beta^{t}\mathbf{T}$ so that the right hand side of (\ref{cocoboule}) is
\begin{align*}
\big(\alpha^{{t}}-\beta^{{t}}\mathbf{T}\big)x-\beta^{{t}}\mathbf{TB}^{{t}}f(\mathbf{A}x)&=a^{t}\mathbf{A}x-\beta^{{t}}\mathbf{TB}^{{t}}f(\mathbf{A}x)\\
&=a^{t}u-\beta^{{t}}\mathbf{TB}^{{t}}f(u) \\ 
&=a^{t}u-b^{t}f(u),
\end{align*}
where we let $u=\mathbf{A}x$ and $b=\beta^{t}\mathbf{T}\mathbf{B}^{{t}}$.
Since $a^{t}\mathbf{A}=\alpha^{t}-\beta^{t}\mathbf{T}$, then by using the canonical right projection for $\mathbf{A}$, we have
$a^{t}\mathbf{A}\mathbf{R_{A}}=a^{t}=\big(\alpha^{t}-\beta^{t}\mathbf{T}\big)\mathbf{R_{A}}$ if and only if $a=\mathbf{R_{A}^{\text{t}}}\big(\alpha-\mathbf{T}^{t}\beta\big)$.
\end{proof}

\subsection{Composition of linear extensions}\label{subsect_comp}

We would like to generalize Lemmas \ref{lemma_fin_diff} and \ref{lemma_lin_approx} for a composition of linearly extended non-linear functions. We recall that $\ell$ functions $F_{k_{i}}$ for $1\leq i\leq \ell$ are composed as $F_{k_{\ell}}\circ\cdots\circ F_{k_{1}}$ where $k_{i}$ denotes the $i$th key. For simplicity and compactness of notation, we denote $F_{k_i}$ by $F_{i}$, and $f_{k_i}$ by $f_{i}$, $1\leq i\leq \ell$. Given a matrix $\mathbf{M}$, we may use $\textrm{im}{\hspace{1pt}\mathbf{M}}$, and $\textrm{colsp}{\hspace{1pt}\mathbf{M}}$ interchangeably.

\begin{theorem}\label{thm_fd}
Let $1<\ell'\leq \ell$, $a\in\bigcap_{j=0}^{\ell'-1}{\mathbf{T}^{j}\ker{A}}$. Then $F_{\ell'}\circ\cdots \circ F_{1}(x+\alpha) - F_{\ell'}\circ\cdots \circ F_{1}(x)=\beta$ has a solution with $\alpha=\mathbf{T}^{-\ell'+1}a$ and $\beta=\mathbf{T}^{\ell'}\alpha$.
\end{theorem}
\begin{proof}
Let $\alpha\in\mathbb{F}^{n}$, and consider the effect of a translation by $\alpha$ on $F_{\ell}\circ\cdots\circ F_{1}(x+\alpha)$. We start with $F_{1}$:
\begin{align*}
F_{1}(x+\alpha)&=\mathbf{T}\big(x+\alpha+\mathbf{B}^{\text{t}}f_{1}(\mathbf{A}(x+\alpha))\big)\\
&=\mathbf{T}\big(x+\mathbf{B}^{t}f_{1}(\mathbf{A}x)\big)+\mathbf{T}\alpha\quad\text{if $\alpha\in\ker{\mathbf{A}}$}\\
&=F_{1}(x)+\mathbf{T}\alpha.\
\end{align*}
Keeping $\alpha\in\ker{\mathbf{A}}$, we proceed to the second iteration:
\begin{align*}
F_{2}\circ F_{1}(x+\alpha)&=F_{2}(F_{1}(x)+\mathbf{T}\alpha)\\
&=\mathbf{T}\big(F_{1}(x)+\mathbf{T}\alpha+\mathbf{B}^{t}f_{2}(\mathbf{A}(F_{1}(x)+\mathbf{T}\alpha))\big)\\
&=\mathbf{T}\big(F_{1}(x)+\mathbf{B}^{t}f_{2}(\mathbf{A}F_{1}(x))\big)+\mathbf{T}^{2}\alpha\quad\text{if $\mathbf{T}\alpha\in\ker{\mathbf{A}}$}\\
&=F_{2}\circ F_{1}(x)+\mathbf{T}^{2}\alpha.
\end{align*}

For some $1<\ell'\leq \ell$, suppose now that $\alpha\in\bigcap_{j=0}^{\ell'-1}{\mathbf{T}^{-j}\ker{\mathbf{A}}}$. We have
\begin{displaymath}
f_{\ell'}\big(\mathbf{A}F_{\ell'-1}\circ\cdots\circ F_{1}(x)+\mathbf{A}\mathbf{T}^{\ell'-1}\alpha\big)=f_{\ell'}(\mathbf{A}F_{\ell'-1}\circ\cdots\circ  F_{1}(x)).
\end{displaymath}
For notational convenience and clarity, we write $y=F_{\ell'-1}\circ\cdots\circ F_{1}(x)\in\mathbb{F}^{n}$ and $a=\mathbf{T}^{\ell'-1}\alpha\in\mathbb{F}^{n}$ so that the latter equation is also rewritten as $f_{\ell'}\big(\mathbf{A}(y+a)\big)=f_{\ell'}(\mathbf{A}y)$. Hence we obtain that
\begin{align}
F_{\ell'}\circ\cdots\circ F_{1}(x+\alpha) &= \mathbf{T}\Big(F_{\ell'-1}\circ\cdots\circ F_{1}(x+\alpha) +\nonumber\\
&\qquad\qquad\qquad \mathbf{B}^{t}f_{\ell'}\big(\mathbf{A}F_{\ell'-1}\circ\cdots\circ F_{1}(x+\alpha)\big)\Big)\label{move_to_nonlinear1}\\
&= \mathbf{T}\Big(F_{\ell'-1}\circ\cdots\circ F_{1}(x) + \mathbf{B}^{t}f_{\ell'}\big(\mathbf{A}(y + a)\big)\Big)+\mathbf{T}a\label{move_to_nonlinear2}\\
&=\mathbf{T}\Big(F_{\ell'-1}\circ\cdots\circ F_{1}(x) + \mathbf{B}^{t}f_{\ell'}(\mathbf{A}y)\Big)+\mathbf{T}a\label{move_to_nonlinear3}\\
&=F_{\ell'}\circ\cdots\circ F_{1}(x)+\mathbf{T}^{\ell'}\alpha.\label{move_to_nonlinear4}
\end{align}
The step from (\ref{move_to_nonlinear1}) to (\ref{move_to_nonlinear2}) is implied by $\alpha\in\bigcap_{j=0}^{\ell'-1}{\mathbf{T}^{-j}\ker{\mathbf{A}}}$ and the definitions of $y$ and $a$. The step from (\ref{move_to_nonlinear2}) to (\ref{move_to_nonlinear3}) is justified since $f_{\ell'}(A(y+a))=f_{\ell'}(Ay)$. The step from (\ref{move_to_nonlinear3}) to (\ref{move_to_nonlinear4}) is implied by the defining Equation (\ref{def_eq_rndfct}) and because $a=\mathbf{T}^{\ell'-1}\alpha$. Thus, Equation (\ref{move_to_nonlinear4}) is equivalent to $F_{\ell'}\circ\cdots\circ F_{1}(x+\alpha)-F_{\ell'}\circ\cdots\circ F_{1}(x)=\beta$ with $\beta=\mathbf{T}^{\ell'}\alpha$.
\end{proof}

In Theorem \ref{thm_lform} that follows, $F_{0}$ is the identity function. We recall also that $\text{im}\hspace{1pt}\mathbf{A} = \ker{\mathbf{B}}$.

We explain now how to refine slightly Theorem \ref{thm_fd}. We recall that $\ell'>1$ and $y=F_{\ell'-1}\circ\cdots\circ F_{1}(x)\in\mathbb{F}^{n}$. Suppose that $a\in\bigcap_{j=0}^{\ell'-2}{\mathbf{T}^{j}\ker{A}}$ \emph{and} $a\notin\bigcap_{j=0}^{\ell'-1}{\mathbf{T}^{j}\ker{A}}$. In addition, suppose there are $b\in\mathbb{F}^{n-m}$, $z\in\mathbb{F}^{m}$ such that $f_{\ell'}(z+\mathbf{A}a)=f_{\ell'}(z)+b$. For a given $(a,b)$, let us denote the set of such $z$'s by $S(a,b)$. If $y\in S(a,b)$ then
\begin{align*}
F_{\ell'}\circ\cdots\circ F_{1}(x+\alpha) &= \mathbf{T}\Big(F_{\ell'-1}\circ\cdots\circ F_{1}(x+\alpha) +\nonumber\\
&\qquad\qquad\qquad \mathbf{B}^{t}f_{\ell'}\big(\mathbf{A}F_{\ell'-1}\circ\cdots\circ F_{1}(x+\alpha)\big)\Big)\\
&=\mathbf{T}\Big(F_{\ell'-1}\circ\cdots\circ F_{1}(x) + \mathbf{B}^{t}f_{\ell'}\big(\mathbf{A}(y + a)\big)\Big)+\mathbf{T}a\\
&=\mathbf{T}\Big(F_{\ell'-1}\circ\cdots\circ F_{1}(x) + \mathbf{B}^{t}\big(f_{\ell'}(\mathbf{A}y)+b\big)\Big)+\mathbf{T}a\\
&=F_{\ell'}\circ\cdots\circ F_{1}(x)+\mathbf{T}^{\ell'}\alpha+\mathbf{T}\mathbf{B}^{t}b\\
&=F_{\ell'}\circ\cdots\circ F_{1}(x)+\beta.
\end{align*}

We point out that it may happen for a given element $a$ that $S(a,b)=\emptyset$ and we therefore cannot reduce the finite difference equation of the linear extension to the smaller non-linear function. We have therefore the following corollary for which its proof follows immediately from the preceding paragraph.
\begin{corollary}
Let $1<\ell'\leq \ell$, $a\in\bigcap_{j=0}^{\ell'-2}{\mathbf{T}^{j}\ker{A}}$ and $a\notin\bigcap_{j=0}^{\ell'-1}{\mathbf{T}^{j}\ker{A}}$.
If there exists $b\in\mathbb{F}^{n-m}$ such that $f_{\ell'}\big(\mathbf{A}(y+a)\big)=f_{\ell'}(\mathbf{A}y)+b$ with $y=F_{\ell'}\circ\cdots \circ F_{1}(x)$, then $F_{\ell'}\circ\cdots \circ F_{1}(x+\alpha) - F_{\ell'}\circ\cdots \circ F_{1}(x)=\beta$ has a solution with $\alpha=\mathbf{T}^{-\ell'+1}a$ and $\beta=\mathbf{T}^{\ell'}\alpha+\mathbf{T}\mathbf{B}^{t}b$.
\end{corollary}

\begin{theorem}\label{thm_lform}
Let $1\leq\ell'<\ell$, $\mathbf{C}\in\mathbb{F}^{n\times m}$ be such that $\mathbf{B}^{t}f_{\ell-\ell'}(y)=\mathbf{C}y$ has a solution. Then the equation $\beta^{t}F_{\ell}\circ \cdots \circ F_{1}(x)-\alpha^{t}F_{\ell-\ell'-1}(x)\circ\cdots\circ F_{1}=0$ has a solution with $\beta^{t}\in\bigcap_{j=1}^{\ell'}{\mathbf{T}^{-j}\mathrm{im}\hspace{1pt}\mathbf{A}}$ and $\alpha^{t}=\beta^{t}\mathbf{T}^{\ell'+1}\big(\mathbf{I}_{n}+\mathbf{C}\mathbf{A}\big)$.
\end{theorem}
\begin{proof}
Let $\beta\in\mathbb{F}^{n}$ and consider the effect of acting over the range of $F_{\ell}\circ\cdots\circ F_{1}$ through linear combinations, that is,
\begin{align*}
\beta^{t}F_{\ell}\circ\cdots\circ F_{1}(x)&=\beta^{t}\big(\mathbf{T}F_{\ell-1}\circ\cdots\circ F_{1}(x)+\mathbf{T}\mathbf{B}^{t}f_{\ell}(\mathbf{A}F_{\ell-1}\circ\cdots\circ F_{1})\big)\\
&=\beta^{t}\mathbf{T}F_{\ell-1}\circ\cdots\circ F_{1}(x)\quad\text{if $\beta^{t}\in\mathbf{T}^{-1}\textrm{im}{\hspace{1pt}\mathbf{A}}$}.
\end{align*}
If $\beta^{t}\in\mathbf{T}^{-1}\textrm{im}\hspace{1pt}\mathbf{A}$, then we proceed to the penultimate iteration:
\begin{align*}
\beta^{t}F_{\ell}\circ\cdots\circ F_{1}(x)&=\beta^{t}\mathbf{T}F_{\ell-1}\circ\cdots\circ F_{1}(x)\\
&=\beta^{t}\mathbf{T}\big(\mathbf{T}\big(F_{\ell-2}\circ\cdots\circ F_{1}(x)+\mathbf{B}^{t}f_{\ell-1}(\mathbf{A}F_{\ell-2}\circ\cdots\circ F_{1}(x))\big)\\
&=\beta^{t}\mathbf{T}^{2}F_{\ell-2}\circ\cdots\circ F_{1}(x) + \beta^{t}\mathbf{T}^{2}\mathbf{B}^{t}f_{\ell-1}(\mathbf{A}F_{\ell-2}\circ\cdots\circ F_{1}(x))\\
&=\beta^{t}\mathbf{T}^{2}F_{\ell-2}\circ\cdots\circ F_{1}(x)\quad\text{if $\beta^{t}\in\mathbf{T}^{-2}\textrm{im}\hspace{1pt}\mathbf{A}$}.
\end{align*}

For some $1\leq\ell'<\ell$, suppose $\beta^{t}\in\bigcap_{j=1}^{\ell'}{\mathbf{T}^{-j}\textrm{im}\hspace{1pt}\mathbf{A}}$ so that
\begin{align}
 & \beta^{t}F_{\ell}\circ\cdots\circ F_{1}(x)=\beta^{t}\mathbf{T}^{\ell'}F_{\ell-\ell'}\circ\cdots\circ F_{1}(x)\nonumber\\
&\hspace{8pt} = \beta^{t}\mathbf{T}^{\ell'+1}\big(F_{\ell-(\ell'+1)}\circ\cdots\circ F_{1}(x) + \mathbf{B}^{t}f_{\ell-\ell'}(\mathbf{A}F_{\ell-(\ell'+1)}\circ\cdots\circ F_{1}(x))\big).\label{black_hole_flare1}
\end{align}

Suppose that we have at our disposal some knowledge about the solution space involving $f_{\ell-\ell'}$ so that we do not proceed any further with the annihilation of the latter. For notational convenience we denote $y=\mathbf{A}F_{\ell-\ell'-1}\circ\cdots\circ F_{1}(x)$. At our disposal is a matrix $\mathbf{C}\in\mathbb{F}^{n\times m}$ such that
\begin{equation}
\mathbf{B}^{t}f_{\ell-\ell'}(y)=\mathbf{C}y.\label{black_hole_flare2}
\end{equation}
Let $b_{i}^{t}\in\mathbb{F}^{1\times (n-m)}$ be the $i$th row of $\mathbf{B}^{t}$ for $1\leq i\leq n$. Equation (\ref{black_hole_flare2}) means there are $a_i^{t}\in\mathbb{F}^{1\times m}$ such that $b_{i}^{t}f_{\ell-\ell'}(y)=a_{i}^{t}y$ where $a_{i}^{t}$ is the $i$th row of $\mathbf{C}$. 

Now recalling Equation (\ref{black_hole_flare1}), we thus have
\begin{align}
&\beta^{t}F_{\ell}\circ\cdots\circ F_{1}(x)=\nonumber\\
&\quad=\beta^{t}\mathbf{T}^{\ell'+1}\big(F_{\ell-(\ell'+1)}\circ\cdots\circ F_{1}(x) + \mathbf{B}^{t}f_{\ell-\ell'}(\mathbf{A}F_{\ell-(\ell'+1)}\circ\cdots\circ F_{1}(x))\big)\nonumber\\
&\quad=\beta^{t}\mathbf{T}^{\ell'+1}\big(F_{\ell-\ell'-1}\circ\cdots\circ F_{1}(x)+\mathbf{B}^{t}f_{\ell-\ell'}(y)\big)\nonumber\\
&\quad=\beta^{t}\mathbf{T}^{\ell'+1}\big(F_{\ell-\ell'-1}\circ\cdots\circ F_{1}(x)+\mathbf{C}y\big)\nonumber\\
&\quad=\beta^{t}\mathbf{T}^{\ell'+1}\big(F_{\ell-\ell'-1}\circ\cdots\circ F_{1}(x)+\mathbf{CA}F_{\ell-\ell'-1}\circ\cdots\circ F_{1}(x)\big)\nonumber\\
&\quad=\beta^{t}\mathbf{T}^{\ell'+1}\big(\mathbf{I}_{n}F_{\ell-\ell'-1}\circ\cdots\circ F_{1}(x)+\mathbf{CA}F_{\ell-\ell'-1}\circ\cdots\circ F_{1}(x)\big)\nonumber\\
&\quad=\beta^{t}\mathbf{T}^{\ell'+1}\big(\mathbf{I}_{n}+\mathbf{C}\mathbf{A}\big)F_{\ell-\ell'-1}\circ\cdots\circ F_{1}(x)\nonumber
\end{align}
Therefore with the choice $\alpha^{t}=\beta^{t}\mathbf{T}^{\ell'+1}\big(\mathbf{I}_{n}+\mathbf{C}\mathbf{A}\big)$, we are able to solve for the equation $\beta^{t}F_{\ell}\circ\cdots\circ F_{1}(x)-\alpha^{t}F_{\ell-\ell'-1}\circ\cdots\circ F_{1}(x)=0$. We finally observe that $\alpha$ depends on $\beta$.
\end{proof}

The following lemma is about the subadditivity property for the dimensions of complementary subspaces of a vector space and can be found for instance in \cite{Godement}. We state it in our language and reprove it for the sake of completeness. We mention Lemma \ref{lem_codim} because it exhibits a relation between the number $\ell$ of iterations and the dimensions $m$ and $n$ of the vector spaces involved in the defining Equation (\ref{def_eq_rndfct}).

\begin{lemma}\label{lem_codim}
With $n$ and $m$ as in the defining Equation (\ref{def_eq_rndfct}), let $\ell>0$ as in (\ref{multiple_iter_def_eq}). Let $0\leq \eta\leq n$ be defined by
\begin{displaymath}
n-\eta=\textrm{codim}\Bigg(\bigcap_{j=0}^{\ell-1}\mathbf{T}^{j}\ker{\mathbf{A}}\Bigg),
\end{displaymath}
where dimensions here are taken over $\mathbb{F}$. Then we have $\ell \geq \frac{n-\eta}{m}$.
\end{lemma}

\begin{proof}
Because $\mathbf{T}$ is invertible, we have for all $j\geq 0$ that
\begin{displaymath}
\dim(\mathbf{T}^{j}\ker{\mathbf{A}})=n-m=\dim(\ker{\mathbf{A}}),
\end{displaymath}
and therefore
\begin{displaymath}
\text{codim}(\mathbf{T}^{j}\ker{\mathbf{A}})=\text{codim}(\ker{\mathbf{A}})=m.
\end{displaymath}
The proof is completed because
\begin{displaymath}
n-\eta=\text{codim}\Bigg(\bigcap_{j=0}^{\ell-1}\mathbf{T}^{j}\ker\mathbf{A}\Bigg)\leq \sum_{j=0}^{\ell-1}{\text{codim}\big(\mathbf{T}^{j}\ker\mathbf{A}\big)} = \sum_{i=0}^{\ell-1}{m}=m\ell.
\end{displaymath}
\end{proof}

To end this section, we make some remarks that we hope are pedagogical.

\begin{remark}
If we replace $\ker{\mathbf{A}}$ by $\ker{\mathbf{B}}$, then Lemma \ref{lem_codim} holds with $m$ replaced by $n-m$ given that $\mathrm{rank}\hspace{1pt}\mathbf{B}=n-m$. Lemma \ref{lem_codim} holds also if $\mathbf{T}$ is replaced by either $\mathbf{T}^{-1}$, $\mathbf{T}^{t}$ or $(\mathbf{T}^{-1})^{t}$. The bounds could very likely be improved as well for certain classes of $\mathbf{T}$.
\end{remark}

\begin{remark}
The $\ker{\mathbf{A}}$ is used in Theorem \ref{thm_fd} to express conditions on $\alpha$ and hence $\beta$. The $\text{im}\hspace{1pt}\mathbf{A}$ is used in \ref{thm_lform} to express conditions on $\beta$ and hence $\alpha$. We end Theorem \ref{thm_fd} by mentioning that $\beta$ depends on our initial choice of $\alpha$. We end Theorem \ref{thm_lform} by mentioning that $\alpha$ depends on our initial choice of $\beta$.
\end{remark}

\begin{remark}
The word ``second'' from Theorem \ref{thm_fd} should be compared with its antonym ``penultimate'' from Theorem \ref{thm_lform}. The interval for $\ell'$ in Theorem \ref{thm_fd} is $1<\ell'\leq \ell$. The interval for $\ell'$ in Theorem \ref{thm_lform} is $1\leq\ell'<\ell$.
\end{remark}

\begin{remark}
For solving a finite difference equation, we proceed in the opposite way than for solving a linear functional equation or vice-versa. For a finite difference equation, the translation occurs in the domain of $F_{\ell}\circ\cdots\circ F_{1}$; therefore we begin the analysis of its effect on $F_{1}$, then on $F_{2}$, and so on until some knowledge about an intermediate non-linear function can be used. For a linear functional equation, then we seek to find a linear correlation (form) on the components of $F_{\ell}\circ\cdots\circ F_{1}$ with the components of an element from its domain; therefore we begin the correlation analysis from $F_{\ell}$, then on $F_{\ell-1}$, and so on until some knowledge about an intermediate non-linear function can be used.
\end{remark}

\section{Further research and conclusion}

We hope that the flexibility of our method eases mathematical analysis of some of the aforementioned cryptographic primitives and paves the way for new ones.

An important question that we would like to answer concerns the strong pseudo-randomness property of the permutation \emph{given} the non-linear functions as defined in \cite{Gol_book2010}. For instance, a composition of four independent Feistel functions, as in Example \ref{ex_feistel}, yields a strongly pseudo-random permutation as shown in \cite{LubRac_1988}. Therefore it strongly suggests that composing a few linear extensions yields a pseudo-random permutation conditional upon pseudo-randomness of the extended non-linear functions. The results from this paper should be useful in proving this pseudo-randomness as well.

Other interesting questions and research lines are the following:
\begin{itemize}
\item[$\bullet$] To establish the cardinalities (or tight bounds) for the solution spaces and their dual spaces with respect to the number of iterations, the dimensions of the vector spaces, the characteristic of the finite field, and the cardinalities for the solution spaces involving the smaller non-linear functions.
\item[$\bullet$] To redo Theorems \ref{thm_fd} and \ref{thm_lform} for second order finite difference equations and for quadratic functional equations.
\item[$\bullet$] If the matrices $\mathbf{A}$ and $\mathbf{B}$ are the generator and parity check matrices of a linear code, then there is no binary linear code that satisfy the Singleton's bound other than the trivial linear codes. Therefore how much would we gain to work over non-binary finite fields in order for $\mathbf{A}$ and $\mathbf{B}$ to satisfy the Singleton's bound property which is also known as the MDS property. What are the implications for $\mathbf{A}$ and $\mathbf{B}$ to be MDS matrices?
\item[$\bullet$] What would be the implications if the construction is changed so that the non-linear part is fixed and we use a random linear code with generator and parity check matrices $\mathbf{A}$ and $\mathbf{B}$?
\item[$\bullet$] For those researchers in applied cryptography, make a catalog of all known ciphers that fit into our scheme or that require slight modifications in order to fit into our scheme.
\end{itemize}
The authors welcome any suggestions or discussions about the previous questions.

\section*{Acknowledgement}

The second author is partially funded by NSERC of Canada. We thank the anonymous 
referees for insightful comments and corrections that improved the paper considerably.

\bibliographystyle{plain}
\newcommand{\SortNoop}[1]{}

\end{document}